\documentclass[11pt]{article}

\usepackage{amsfonts,amssymb}
\usepackage{amssymb}
\begin{document}

    \makeatletter \@addtoreset{equation}{section} \makeatother
    \renewcommand{\theequation}{\thesection.\arabic{equation}}
    \baselineskip 15pt

\newtheorem{defofentangidentical}{Definition}[section]
\newtheorem{defofentangidentical2}[defofentangidentical]{Definition}
\newtheorem{factorizabilityidentical2}{Theorem}[section]
\newtheorem{factorizabilityidentical3}[factorizabilityidentical2]{Theorem}
\newtheorem{anticomplex}{Theorem}[subsection]
\newtheorem{fermionschmidt}[anticomplex]{Theorem}
\newtheorem{measurefermion}[anticomplex]{Theorem}
\newtheorem{simcomplex}{Theorem}[subsection]
\newtheorem{bosonschmidt}[simcomplex]{Theorem}
\newtheorem{measureboson}[simcomplex]{Theorem}
\newtheorem{bequalb}[simcomplex]{Theorem}
\newtheorem{bdifferentb}[simcomplex]{Theorem}
\newtheorem{entropyin01}[simcomplex]{Theorem}

\title{\bf  General  criterion  for  the  entanglement  of  two
indistinguishable  particles\footnote{Work  supported  in part  by
Istituto Nazionale di Fisica Nucleare, Sezione di Trieste, Italy}}

\author{GianCarlo Ghirardi\footnote{e-mail: ghirardi@ts.infn.it}\\
{\small Department of Theoretical Physics of the University of
Trieste, and}\\ {\small International Centre
for Theoretical Physics ``Abdus Salam'', and}\\ {\small Istituto
Nazionale di Fisica Nucleare, Sezione di Trieste, Trieste, Italy}\\ and \\
\\ Luca Marinatto\footnote{e-mail: marinatto@ts.infn.it}\\ {\small
Department of Theoretical Physics of the University of Trieste,
and}\\ {\small Istituto Nazionale di Fisica
Nucleare, Sezione di Trieste, Trieste, Italy}}

\date{}

\maketitle
\begin{abstract}
We relate the notion of entanglement for quantum
 systems composed of two identical constituents to the impossibility of
 attributing a complete set of properties to both particles.
This  implies definite constraints on the mathematical form of the
 state vector associated with the whole system.
We then analyze separately the  cases of fermion and boson systems,
 and we show how the consideration of {\em both} the Slater-Schmidt number of 
 the fermionic and bosonic analog of the Schmidt decomposition of
 the global state vector {\em and} the von Neumann entropy of the one-particle
 reduced density operators can supply us with a consistent criterion
 for detecting entanglement. In particular, the consideration of the
 von Neumann entropy is particularly useful in deciding
 whether the correlations of the considered states are simply due to
 the indistinguishability of the particles involved or are a genuine
 manifestation of the entanglement.
The treatment leads to a full clarification of the subtle aspects of
 entanglement of two identical constituents which have
 been a source of embarrassment and of serious misunderstandings in
 the recent literature.\\



\end{abstract}


\section{Introduction}

Quantum entanglement, considered by Schr\"odinger
 {\em ``the characteristic trait of Quantum Mechanics, the one that
 enforces its entire departure from classical lines of thoughts''}~\cite{sch},
 has played a central role in the historical development of Quantum
 Mechanics and  nowadays it constitutes  an essential resource for many
 aspects of quantum information and quantum computation theory.
In fact, the possibility of performing reliable teleportation processes,
 of generating unconditionally secure private keys in cryptography,
 or of devising quantum algorithms allowing us to solve certain computational
 problems in a more efficient way than the best known classical methods,
 is essentially based on the peculiar properties of entangled states.
However, in spite  of the fact that entangled states {\it involving
 identical constituents} are widely used in the experimental
 implementations of the above mentioned (and many other) processes, 
 the very notion of entanglement for such ubiquitous physical systems seems 
 to be too often misunderstood, or not understood at all, in the current 
 scientific literature on the subject.
The most frequent misinterpretations arise in connection with the
 symmetrization postulate of quantum mechanics which requires definite
 symmetry properties for the state vectors associated with systems of
 identical particles.
Their nonfactorized form seems to suggest~\footnote{The only important 
 exception being represented by the peculiar case of identical bosons in
 the same state.}, when compared with the well-known
 case of systems composed of distinguishable particles, the occurrence of
 an unavoidable form of  entanglement when  identical particles enter
 into play.

In a recent paper~\cite{gmw} (see also \cite{gm}) we analyzed in
 great detail the problem of entanglement, dealing with systems of
 two (or more) both distinguishable and identical particles. In accordance with
 the position of the founding fathers of quantum mechanics~\cite{sch,epr},
 we strictly related the nonoccurrence of entanglement to the
 possibility of attributing {\em complete sets of properties} with both
 constituents of the composite system.
In this way we were able to formulate an unambiguous criterion for
 deciding whether a given state vector is entangled or not, which works
 for the cases of both distinguishable and identical constituents.
It has to be stressed that, contrary to what has sometimes been stated,
 non-entangled states involving identical constituents can actually occur.

Obviously, in the case of distinguishable particles, our criterion is
 equivalent (as we showed in Ref.~\cite{gmw}) to the  commonly
 used criteria to identify whether a system is entangled or not, which
 involve consideration of the Schmidt number of the biorthonormal
 decomposition of the state or, equivalently, the evaluation of the
 von Neumann entropy of the reduced statistical operators.

The situation is radically different in the case of composite systems
 involving identical constituents.
In the literature, various authors~\cite{cirac,pask,li} have suggested
 identifying the entangled or non-entangled nature of a system of two
 identical constituents by resorting to natural generalizations of
 the above mentioned criteria.
In so doing they have met various difficulties which emerge when one
 compares the results obtained with the analogous ones for the case of
 distinguishable particles.
Moreover, the procedure yields apparently contradictory results for
 the fermion and boson cases~\cite{pask,li,vacca}.

In this paper, we show that the source of the problems rests in not
 having appropriately taken into account the fact that, even in the
 case in which it is physically legitimate and
 correct to consider a state as non-entangled, so that one knows the
 properties of the constituents, there is an unavoidable lack of
 information about the actual situation of the constituents,
 arising from their identity.
Furthermore, we prove that by resorting to the general analysis of
 Refs.~\cite{gmw,gm} one can get a complete clarification of the
 matter and we present a unified criterion for detecting entanglement in
 systems of identical particles such that (i) it involves {\em both} the
 Slater-Schmidt number of the fermionic and bosonic analog of the Schmidt
 decomposition of the state vectors {\em and} the von Neumann entropy of
 the reduced single-particle statistical operators; (ii) it applies equally
 well for fermions and bosons; and (iii) it is in complete accordance with our
 original criterion.
     

\section{Entanglement and properties}

In this section we briefly review the arguments of Ref.~\cite{gmw},
 which show that the correct way  to identify the entanglement is 
 to relate it to the
 impossibility of attributing precise properties to the (identical)
 constituents of a two-particle system $S=S_{1}+S_{2}$.
As is well known, in the case of non-entangled distinguishable particles
 the factorized nature of the state vector
 $\vert\psi(1,2)\rangle=\vert \phi \rangle_{1} \otimes \vert\chi
 \rangle_{2}$ is a necessary and sufficient condition for being allowed to 
 claim that subsystem $S_{1}$ objectively
 possesses the (complete set of) properties associated with the state
 vector $\vert \phi \rangle_{1}$ and subsystem $S_{2}$ those associated
 with $\vert\chi \rangle_{2}$.
We stress that in the case considered we know not only the properties
 that are possessed but also to which system they refer. Obviously, in the 
 case of identical constituents one cannot resort to the factorizability
 criterion to claim that the two systems are non-entangled, otherwise one would
 be
 led to conclude (mistakenly) that non-entangled states
 cannot exist (an exception being made for two bosons in the same state), since
 the necessary symmetry requirements forbid the occurrence of factorized
 states.

However, this naive and inappropriate conclusion derives from taking
 a purely formal attitude about the problem, without paying due
 attention to the physically meaningful conditions which, when satisfied,
 allow one to legitimately state that two systems are non-entangled.
Such conditions are, primarily, those of being
 allowed to claim that one particle possesses a precise and
 complete set of properties and the other one exhibits analogous features.
Obviously, one must always keep clearly in mind that it is absurd to
 pretend to individuate the particles, i.e., to identify which one possesses 
 one set of properties and which one the other set (in the case in which such 
 sets are different).

Considerations of this kind have led us to identify the following
 physically appropriate criterion characterizing non-entangled states
 of two identical particles.
\begin{defofentangidentical}
 \label{defofentangidentical}
 The identical constituents ${\cal S}_{1}$ and ${\cal S}_{2}$ of a
 composite quantum system ${\cal S}={\cal S}_{1}+{\cal S}_{2}$ are {\bf
 non-entangled} when both constituents possess a complete set of properties.
\end{defofentangidentical}
\noindent Obviously, we still have to make fully precise the meaning
 of the expression {\em ``both constituents possess a complete set of
 properties"}.
To this end we resort, first of all, to the following definition:
\begin{defofentangidentical2}
 \label{defofentangidentical2}
 Given a composite quantum system ${\cal S}={\cal S}_{1}+{\cal S}_{2}$ of
 two identical particles described by the normalized state vector
 $\vert \psi(1,2) \rangle$, we will say that one of the constituents
 possesses a complete set
 of properties if and only if there exists a one-dimensional projection 
 operator $P$, defined on the single particle Hilbert space ${\cal H}$, 
 such that:
\begin{equation}
 \label{wittgenstein}
 \langle \psi(1,2)\vert\,{\cal E}_{P}(1,2)\,\vert \psi(1,2)\rangle =1
\end{equation}
\noindent where
\begin{equation}
 \label{operatorediproiezione}
 {\cal E}_{P}(1,2)=P^{(1)}\otimes [\,I^{(2)}-P^{(2)}\,] +
 [\,I^{(1)}-P^{(1)}\,]\otimes  P^{(2)} + P^{(1)}\otimes P^{(2)}.
\end{equation}
\end{defofentangidentical2}
Condition~(\ref{wittgenstein}) gives the probability of finding {\it at
 least} one of the two identical particles (of course we cannot say
 which one) in the state associated with the
 one-dimensional projection operator $P$~\footnote{We remark that one could
 drop the last term in the expression of Eq.~(\ref{operatorediproiezione}),
 getting an operator whose
 expectation value gives the probability of finding {\it precisely}
 one particle in the state onto which $P$ projects. In the case of
 identical fermions this makes no difference,
 but for bosons it would not cover the case of both particles being in
 the same state.}.
Since any state vector is a simultaneous eigenvector of a complete set of
 commuting observables, condition~(\ref{wittgenstein}) allows us to attribute
 to at least one of the particles the complete set of properties
 (eigenvalues) associated with the considered set of observables.

At this point we must distinguish two cases. If the two identical
 particles are fermions, then one can immediately prove (see Ref.~\cite{gmw})
 that Eq.~(\ref{wittgenstein}) implies that there exists another
 one-dimensional projection operator $Q$, which is orthogonal to $P$, such
 that the operator ${\cal E}_{Q}$,
 which has the expression~(\ref{operatorediproiezione}) with $Q$ replacing
 $P$, also satisfies
\begin{equation}
 \label{wittgenstein2}
 \langle \psi(1,2)\vert\,{\cal E}_{Q}(1,2)\,\vert \psi(1,2)\rangle =1.
\end{equation}
Obviously, in such a case, since it is simultaneously true that
 ``{\it there is at least one particle having the properties
 associated with P}" and ``{\it there is at least one particle
 having the properties associated with Q}" and, moreover, such
 properties are mutually exclusive due to the orthogonality of the
 projection operators, we can legitimately claim
 that, in the state $\vert\psi(1,2)\rangle$,{\em ``one particle (it is
 meaningless to ask which one) has the complete set of properties
 associated with $P$ and one the complete set
 associated with $Q$''}.
As we showed in Ref.~\cite{gmw} our request for non-entanglement
 leads, in the fermion case, to the following theorem:
\begin{factorizabilityidentical2}
 \label{factorizabilityidentical2}
 The identical fermions ${\cal S}_{1}$ and ${\cal S}_{2}$ of a composite
 quantum system ${\cal S}={\cal S}_{1}+{\cal S}_{2}$ described by the pure
 normalized state $\vert \psi(1,2) \rangle$ are {\bf non-entangled} if and
 only if $\vert \psi(1,2) \rangle$ is obtained by antisymmetrizing a factorized
 state.
\end{factorizabilityidentical2}

In the boson case the situation is slightly different since the two 
 particles can be in the same state.
To be completely general let us begin by assuming that at least one of the
 constituents possesses a complete set of properties, so that there 
 exists a one-dimensional single-particle projection operator $P$ such 
 that the associated ${\cal E}_{P}$ satisfies Eq.~(\ref{wittgenstein}).
At this point we take into account the operator $P^{(1)}\otimes P^{(2)}$ and 
 we consider its expectation value in the state $\vert\psi(1,2)\rangle$.
Three possible cases can occur.
\begin{itemize}
\item If $\langle\psi(1,2)\vert P^{(1)}\otimes P^{(2)}\vert\psi(1,2)\rangle=1$,
 then we can say that {\em ``both particles possess the complete set of
 properties associated with $P$''} and  $\vert\psi(1,2)\rangle$ is the tensor
 product of two identical state vectors.
This is the only situation in which the unavoidable ambiguities ensuing from
 the identity of the constituents disappear.
\item  If $\langle\psi(1,2)\vert P^{(1)}\otimes P^{(2)}\vert\psi(1,2)\rangle=0$
 then the condition Eq.(\ref{wittgenstein}) implies that the state is
 obtained by symmetrizing the product of two orthogonal states, one of which is
 the one on which $P$ projects, and, consequently, there is another
 one-dimensional projection operator $Q$ orthogonal to $P$ such that
 Eq.~(\ref{wittgenstein2}) is satisfied, and the mean value of
 $Q^{(1)}\otimes Q^{(2)}$ also vanishes.
The situation is perfectly analogous to the fermion case, and
 one can conclude that {\em ``there is precisely
 one particle possessing the properties identified by $P$ and one possessing
 the properties identified by $Q$''}.
\item If $\langle\psi(1,2)\vert P^{(1)}\otimes P^{(2)}\vert\psi(1,2)\rangle
 \in (0,1)$, since condition Eq.~(\ref{wittgenstein}) implies that the
 state is obtained by symmetrizing the product of two states, one
 of which is the one on which $P$ projects, the second state cannot be 
 orthogonal to the first one.
If we denote as $Q$ the one-dimensional projection operator on such a state,
 then we can immediately verify that Eq.~(\ref{wittgenstein2}) holds and
 that $\langle\psi(1,2)\vert Q^{(1)}\otimes Q^{(2)}\vert\psi(1,2)\rangle
 \in (0,1)$. 
In such a situation, in spite of the fact that the two statements {\em ``there
 is at least one particle with the properties associated with $P$''} and {\em 
 ``there is at least one particle with the properties associated with $Q$''} 
 are true, one cannot conclude that {\em ``there is precisely one particle
 possessing the properties identified by $P$ and one possessing the properties
 identified by $Q$''}.
\end{itemize}
The above considerations, as the reader can easily grasp and as
 has been proved in Ref.~\cite{gmw}, lead to the following theorem:
\begin{factorizabilityidentical3}
 \label{factorizabilityidentical3}
 The identical bosons of a composite quantum system ${\cal S}={\cal S}_{1}
 + {\cal S}_{2}$ described by the pure normalized state $\vert \psi(1,2)
 \rangle$ are {\bf non-entangled} if and only if either the state is obtained 
 by symmetrizing a factorized product of two orthogonal states or it is
 the product of the same state for the two particles.
\end{factorizabilityidentical3}

Concluding, we have shown that, when one deals with the problem of
 entanglement using the necessary logical rigour and making appeal
 to the physical meaning of entanglement itself, then the process of
 (anti)symmetrizing a state vector does not necessarily lead to an
 entangled state.

In addition to the physical motivations we have presented, there are other
 compelling reasons which show that our criterion for non-entanglement
 is the correct one.
It is in fact easy to show that, when the conditions of
 Theorems~\ref{factorizabilityidentical2} or~\ref{factorizabilityidentical3}
 are satisfied, it is not possible to take advantage of the form of
 the state vector to perform teleportation processes or to violate Bell's
 inequality.
These facts further strengthen our conclusion that the state is
 non-entangled.

To clarify our argument we resort to an extremely simple example.
Let us consider the following state of two identical spin-$1/2$ fermions,
 in which we have denoted as $\vert R\rangle$ and $\vert L \rangle$ two
 precise orthonormal states of the single-particle configuration space
 having disjoint supports in two distant and nonoverlapping spatial regions
 Right and Left, respectively, and
 with $\vert z\uparrow \rangle$ and $\vert z\downarrow \rangle$ the
 eigenvectors of the spin observable $\sigma_{z}$:
\begin{equation}
 \label{falsoepr}
 \vert \psi(1,2)\rangle = \frac{1}{\sqrt{2}}\, [\, \vert z \uparrow
 \rangle_{1} \vert R \rangle_{1} \otimes \vert z \downarrow \rangle_{2}
 \vert L\rangle_{2} - \vert z \downarrow \rangle_{1} \vert L \rangle_{1}
 \otimes \vert z \uparrow \rangle_{2}\vert R \rangle_{2} \,].
\end{equation}
Such a state, which can be obtained by antisymmetrizing a factorized
 state (that is, $\vert z \uparrow \rangle_{1} \vert R \rangle_{1} 
 \otimes \vert z \downarrow
 \rangle_{2} \vert L\rangle_{2}$), makes perfectly legitimate a
 statement of the type~\footnote{Note that in making our statement 
 we somehow individuate our  constituents by making reference to the 
 fact that they lie in different spatial regions. This remark is  useful for 
 the analysis that will follow.}: {\em ``there is one particle at the right 
 having spin
 up along the z-direction and the spatial properties associated with
 $\vert R\rangle$ and a particle at the left having spin down along the
 z-direction and the spatial properties associated with $\vert L\rangle$"}.
As a consequence, this state does not imply any (nonlocal) correlation
 between spin measurements performed in the two spacelike separated
 regions Right and Left and thus it cannot violate  Bell's inequality.
This follows immediately from the fact that, when the state is the one of
 Eq.~(\ref{falsoepr}), we have for the mean value $E(\vec{a}, \vec{b})$
 of the product of the outcomes of two spin measurements along the
 directions $\vec{a}$ and $\vec{b}$ in the two regions Right and Left,
 the following expression:
\begin{eqnarray}
\label{falsoepr2}
 E(\vec{a},\vec{b})& = &\langle \psi \vert\:\Big[\, \vec{\sigma}^{(1)}\cdot
 \vec{a} P_{R}^{(1)} \otimes \vec{\sigma}^{(2)}\cdot \vec{b} P_{L}^{(2)} +
 \vec{\sigma}^{(1)}\cdot \vec{b} P_{L}^{(1)}\otimes
 \vec{\sigma}^{(2)}\cdot \vec{a} P_{R}^{(2)}\,\Big]\:\vert \psi \rangle 
 \nonumber \\
&=& \langle z \uparrow \vert
 \vec{\sigma}\cdot \vec{a} \vert z \uparrow \rangle \langle z \downarrow
 \vert \vec{ \sigma}\cdot \vec{b} \vert z \downarrow \rangle\:,
\end{eqnarray}
where we have denoted as $P_{R}$ and $P_{L}$ the projection operators on
 the closed linear manifolds of the spatial wave functions with compact
 support in the Right and Left regions, respectively.
The occurrence of a factorized product of two mean values implies that
 no choice of the unit vectors $\vec{a}$, $\vec{b}$, $\vec{c}$, and
 $\vec{d}$ can lead to a violation of Bell's inequality~\cite{ghsz}:
\begin{equation}
 \vert E(\vec{a},\vec{b}) - E(\vec{a},\vec{c})\vert +
 \vert E(\vec{b},\vec{d}) + E(\vec{c},\vec{d}) \vert  \leq 2\:.
\end{equation}
On the other hand, let us consider a state of the form:
\begin{equation}
 \label{veroepr}
 \vert \phi(1,2)\rangle = \frac{1}{2}\, [\, \vert z \uparrow \rangle_{1}
 \vert z \downarrow \rangle_{2} - \vert z \downarrow \rangle_{1}
 \vert z \uparrow \rangle_{2}\,] \otimes [\, \vert R \rangle_{1}
 \vert L \rangle_{2}+ \vert L \rangle_{1}\vert R \rangle_{2}\,],
\end{equation}
which is the one used in the Bohm version of the Einstein-Podolsky-Rosen
 argument of incompleteness.
Since it cannot be obtained by antisymmetrizing a factorized product of
 two orthogonal states it is a genuine entangled state according to our 
 criterion. Correspondingly, one can easily prove that such a state
 exhibits the nonlocal features leading to a violation of Bell's
 inequality for a proper choice of the orientations
 $\vec{a}$, $\vec{b}$, $\vec{c}$, and $\vec{d}$.

In spite of their simplicity, the examples just considered show
 clearly why vectors displaying the apparent form of an entangled state,
 such as the one of Eq.~(\ref{falsoepr}), must be considered as
 non-entangled, in perfect agreement with our criterion.


\section{The entanglement criteria}

As we have already stressed, when one deals with quantum systems
 composed of two distinguishable particles,
 the appearance of entanglement is equivalent to the impossibility of
 writing the state vector of the compound system $\vert \psi(1,2) \rangle$
 as a tensor product of two single-particle states.
This in turn implies two well-known formal facts: (i) by resorting to the
 Schmidt decomposition, the global state turns out to be non-entangled
 if and only if its associated Schmidt number (that is, the number of
 non-zero coefficients in such a decomposition)  equals  1;
 (ii) the state is non-entangled if and only if the von Neumann entropy of 
 the reduced statistical operator associated with both particles is equal 
 to zero~\footnote{As is well known, given a statistical operator $\rho$, 
 its von Neumann entropy is defined as $S(\rho)\equiv 
 - {\textrm {Tr}}[\,\rho \ln \rho\,]$ where the base of the logarithm 
 function is
 the number $e$. However, in the present paper, we will rescale this
 quantity and follow the information theory convention of using all logarithms
 in base $2$.}. 

Both facts have clear physical implications and a precise meaning:
 the first refers to the possibility of attributing a complete and
 precise set of properties to each constituent, the second ensures that
 we have the most complete and exhaustive information allowed by quantum
 theory about the situation of each constituent.
In fact, in a factorized state each component subsystem is associated with a
 precise state vector and the reduced statistical operator for one
 of the two particles, for example, the one labeled by $1$, i.e.,
 $\rho^{(1)}={\textrm {Tr}}^{(2)}[\,\vert \psi(1,2)\rangle \langle \psi(1,2)
 \vert\,]$, turns out to be  a projection operator onto a one-dimensional 
 manifold.
Correspondingly, its von Neumann entropy $S(\rho^{(1)})\equiv - 
 {\textrm {Tr}}^{(1)}[\, \rho^{(1)}\log \rho^{(1)}\,]$ equals zero.
This result is correct since such a quantity measures the lack of
 information about the single-particle subsystem and there is, in fact, no
 uncertainty at all concerning the state that must be attributed to it.

When passing to the more subtle case of interest, that is, to systems
 composed of two identical constituents, the relations between entanglement 
 and both the Schmidt number and the von Neumann entropy of the reduced
 statistical operators become less clear and require a careful analysis.
The purpose of this Section is to clarify the matter and to 
 present a criterion for determining whether a state is entangled or not 
 which is (i) based  on a consideration of {\em both} the Slater-Schmidt
 number {\em and} the von Neumann entropy; (ii) consistent with our original
 criterion summarized in Definition~\ref{defofentangidentical}; (iii)
 equally applicable to fermion and boson systems; and finally (iv) able to
 unify in a consistent way various criteria which have appeared recently in the
 literature~\cite{cirac,pask,li}.

We  limit our considerations to the case of a finite-dimensional 
 single-particle Hilbert space and, in accordance with the above 
 remarks, we deal separately with the fermion and the boson cases, since 
 they exhibit quite different features.


\subsection{The fermion case}

The notion of entanglement for systems composed of two identical
 fermions has been discussed in Ref.~\cite{cirac} where a 
 {\em ``fermionic analog of the Schmidt decomposition''} was exhibited.
Such a decomposition is based on a nice extension to the set of the
 antisymmetric complex matrices of a well-known theorem holding
 for antisymmetric real matrices (see, for example,~\cite{mehta}).
The Theorem of~\cite{cirac} states the following:
\begin{anticomplex}
 \label{anticomplex}
 For any antisymmetric $(N\times N)$ complex matrix A [that is, $A\in
 {\cal M} (N,\mathbb{C})$ and $A^{T}=-A$] there exists a unitary 
 transformation 
 $U$ such that $A=UZU^{T}$, with $Z$  a block-diagonal matrix of the sort
\begin{equation}
 \label{fermion1}
 Z=diag[\,Z_{0},Z_{1},\dots,Z_{M}\,], \hspace{0.9cm} Z_{0}=0\,,\hspace{0.9cm}
 Z_{i}= \left[
 \begin{array}{cc}
 0 & z_{i}\\
 -z_{i} & 0
 \end{array}
 \right],
\end{equation}
where $Z_{0}$ is the $(N-2M)\times (N-2M)$ null matrix and $z_{i}$ are
 complex numbers.
Equivalently, $Z$ is the direct sum of the $(N-2M)\times 
 (N-2M)$ null matrix and the $M$ $(2\times 2)$ complex antisymmetric matrices 
 $Z_{i}$.
\end{anticomplex}
The fermionic analog of the Schmidt decomposition follows from an application
 of Theorem~\ref{anticomplex} to systems composed of identical fermions:
\begin{fermionschmidt}
 \label{fermionschmidt}
 Any state vector $\vert \psi(1,2)
 \rangle$ describing two identical fermions of spin $s$ and,
 consequently, belonging to the antisymmetric manifold
 ${\cal A}(\mathbb{C}^{2s+1}\otimes \mathbb{C}^{2s+1})$, can be written as:
\begin{equation}
 \label{fermion2}
 \vert \psi(1,2)\rangle = \sum_{i=1}^{(2s+1)/2} a_{i}\cdot
 \frac{1}{\sqrt{2}} \,[\,\vert 2i-1\rangle_{1} \otimes \vert 2i
 \rangle_{2}- \vert 2i \rangle_{1} \otimes \vert 2i-1 \rangle_{2}\,],
\end{equation}
where the states $\left\{\, \vert 2i-1 \rangle, \vert 2i \rangle \right\}$
 with $i=1,\dots, (2s+1)/2$ constitute an orthonormal basis of 
 $\mathbb{C}^{2s+1}$, and the complex coefficients $a_{i}$ (some of which may 
 vanish) satisfy the normalization condition $\sum_{i} \vert a_{i}\vert^{2}=
 1$.
\end{fermionschmidt}

Following the authors of Ref.~\cite{cirac}, the number of nonzero
 coefficients $a_{i}$ appearing in the decomposition~(\ref{fermion2}) is
 called the {\em Slater number} of $\vert \psi(1,2)\rangle$.
The relation of such a number to the notion of entanglement has been
 made explicit in the papers~\cite{pask,li}, where a state
 displaying the form of Eq.~(\ref{fermion2}) is called entangled if and
 only if its Slater number is strictly greater than $1$.
It is worth noticing that this condition turns out to be totally
 equivalent to our Definition~\ref{defofentangidentical}.
In fact, given an arbitrary two-particle state $\vert \psi(1,2) \rangle$,
 suppose that its fermionic Schmidt decomposition has Slater number equal
 to $1$.
According to Theorem~\ref{fermionschmidt}, this  means that
 there exist two orthonormal vectors $\vert 1 \rangle$ and
 $\vert 2 \rangle$ belonging to $\mathbb{C}^{2s+1}$ such that:
\begin{equation}
 \label{fermion3}
 \vert \psi(1,2)\rangle = \frac{1}{\sqrt{2}}\,[\,\vert 1\rangle_{1}\otimes
 \vert 2\rangle_{2} - \vert 2\rangle_{1}\otimes \vert 1\rangle_{2}\,],
 \hspace{1cm} \langle 1 \vert 2 \rangle=0.
\end{equation}
Since the state can be obtained by antisymmetrizing the product state
 $\vert 1 \rangle_{1}\otimes \vert 2\rangle_{2}$,
 the state must be considered as non-entangled in accordance
 with our criterion.
Vice versa, any state obtained by antisymmetrizing a factorized state has
 Slater number equal to 1.
On the contrary, if the Slater number is greater (or equal to) two,
 the form Eq.~(\ref{fermion2}) of the state
 shows immediately that it cannot be obtained by antisymmetrizing a
 product of two orthonormal vectors, so that the state must be considered 
 as a genuinely entangled one.

So far the criterion of Ref.~\cite{cirac,pask,li} and ours agree
 completely; however, some problems arise when one calculates the von 
 Neumann entropy of the reduced density operator associated with one of the 
 two particles~\footnote{Since
 the state vector of the compound system $\vert \psi(1,2)\rangle$ is
 antisymmetric under the exchange of the two particles, its associated
 density operator $\rho^{(1+2)} \equiv \vert \psi(1,2) \rangle \langle
 \psi(1,2)
 \vert$ is a symmetric operator and the reduced density operators of the
 two particles are equal, that is $\rho^{(1)} \equiv 
 Tr^{(2)}[\, \rho^{(1+2)} \,]= \rho^{(2)} \equiv
 Tr^{(1)}[\, \rho^{(1+2)}\,]$.}.
In fact from Eq.~(\ref{fermion2}) one gets:
\begin{equation}
 \label{fermion4}
 \rho^{(1)} \equiv Tr^{(2)}[\, \vert \psi(1,2) \rangle \langle \psi(1,2)
 \vert\,]= \sum_{i} \frac{\vert a_{i}\vert^{2}}{2}\cdot [\,
 \vert 2i-1 \rangle_{1}\langle 2i-1 \vert + \vert 2i \rangle_{1}
 \langle 2i \vert\,]\,.
\end{equation}
The von Neumann entropy for such an operator, which is already  in its
 diagonal form, can be easily calculated:
\begin{equation}
 \label{fermion5}
 S(\rho^{(1)}) \equiv - Tr^{(1)}[\, \rho^{(1)}\log \rho^{(1)}\,] =
 -\sum_{i} \vert a_{i}\vert^{2} \log \frac{\vert a_{i}\vert^{2}}{2}= 1
 - \sum_{i} \vert a_{i}\vert^{2} \log \vert a_{i}\vert^{2}\:.
\end{equation}
It follows trivially that:
\begin{equation}
 \label{fermion5.5}
 S(\rho^{(1)})\geq 1 \hspace{1cm} \forall \:\vert a_{i}\vert^{2}
 \in [0,1] \hspace{0.7cm} \sum_{i} \vert a_{i}\vert^{2}=1\:.
\end{equation}
In analogy with the case of distinguishable particles, one could be
 tempted to regard this quantity as a measure of entanglement.
But, according to the authors of Ref.~\cite{pask} and~\cite{vacca}, this
 naturally raises the following two puzzling issues:
 (i) $S(\rho^{(1)})$ of Eq.~(\ref{fermion5}) attains its minimum
 value $S_{min}= 1$ in correspondence with a state with Slater number equal
 to $1$ (which, in accordance with their position, is assumed to identify a
 non-entangled state), contrary to what happens for a (non-entangled) state 
 of distinguishable particles with Schmidt number equal to $1$, for which the
 value of the entropy still takes its minimum value which, however, equals $0$;
 (ii) moreover, in the case of two identical bosons, the minimum of the
 analogous quantity, as we will show later, is null.
This seems to imply that the von Neumann entropy is inappropriate to deal with
 our problem since it gives different measures of entanglement for boson and
 fermion states.

The problematic aspects of this situation derive entirely from not
 having taken correctly into account the real meaning
 of the von Neumann entropy as a measure of the uncertainty about the
 state of a quantum system.
To be  more precise, let us consider a state $\vert \psi(1,2) \rangle$
 with Slater number equal to $1$, i.e., a non-entangled state according to 
 our and the
 Slater number criterion, like that of Eq.~(\ref{fermion3}).
As we stressed in Section $1$, in this situation we can attribute
 definite quantum states $\vert 1 \rangle$ and $\vert 2 \rangle$ to
 the particles but, since they are totally indistinguishable, we cannot
 know whether, for example, particle $1$ is associated
 to the state $\vert 1\rangle$ or to the state $\vert 2 \rangle$.
As a natural consequence, the reduced density operator of each particle
 and its associated von Neumann entropy reflect such an unavoidable
 ignorance~\footnote{To avoid any misunderstanding we stress that with the
 expression {\em ``ignorance''} we do not intend any lack of knowledge
 about the system. In fact in the considered case this ignorance derives
 entirely from the fact that, within Quantum Mechanics, there is no 
 conceivable way to individuate identical particles.}.
Actually, they are equal to:
\begin{equation}
 \label{fermion6}
 \rho^{(1\:or\:2)} = \frac{1}{2}\,[\, \vert 1 \rangle \langle 1
 \vert + \vert 2 \rangle\langle 2 \vert \,] \hspace{1cm}
 \Rightarrow \hspace{1cm} S(\rho^{(1\:or\:2)})=1.
\end{equation}
It should be obvious that we cannot pretend that the operator
 $\rho^{(1\:or\:2)}$ of Eq.~(\ref{fermion6}) describes the properties of 
 {\em precisely} the
 first or of the second particle of the system: once again, due to the
 subtle implications of the identity in quantum
 mechanics, such an operator describes correctly the properties of a
 randomly chosen particle (a particle that cannot be better {\em identified}).
Accordingly, in this case, the quantity $S(\rho^{(1\:or\:2)})=1$
 correctly measures {\em the uncertainty concerning the quantum state to
 attribute to each of the two identical physical subsystems}, and,
 in this situation, it cannot be
 regarded as a measure of the entanglement of the whole state.

A counterpart of this is the fact that the only quantum correlations
 exhibited by the state $\vert\psi(1,2)\rangle$ of Eq.~(\ref{fermion3})
 are those related to the exchange  properties of the indistinguishable
 fermions.
As we have stressed before, such correlations cannot be used to violate
 Bell's inequality or to perform any teleportation process.
Accordingly, they cannot be considered as a manifestation of
 entanglement.

Continuing our analysis, if we consider a state vector with Slater
 number strictly greater than $1$, i.e., an entangled state according to our 
 and to the Slater number criterion, the von Neumann entropy of the associated
 reduced density operator turns 
 out to be strictly greater than $1$ [see Eq.~(\ref{fermion5.5})].
In addition to the minimum amount of uncertainty deriving from the
 indistinguishability of the particles, there is now the {\em additional}
 ignorance naturally connected with the genuine entanglement of the state.
In this case, in fact, we cannot identify two  quantum states 
 that can be attributed~\footnote{With this expression we mean that a 
 measurement aimed to test whether there is one particle in state 
 $\vert 1\rangle$ and one in state $\vert 2\rangle$ gives with certainty a 
 positive outcome.} to the pair of particles (contrary to the previously 
 described situation).

To conclude this subsection, we  exhibit the criteria for detecting 
 entanglement involving the Slater number or the von Neumann entropy,
 in the case of two identical fermions. 
Such criteria are totally consistent with our original requirement that 
 definite properties can be attributed to both component subsystems of a 
 non-entangled state:
\begin{measurefermion}
 \label{measurefermion}
 A state vector $\vert \psi(1,2) \rangle$ describing two identical
 fermions is {\bf non-entangled} if and only if
 its Slater number is equal to 1 or
 equivalently iff the von Neumann entropy of the one-particle reduced
 density operator $S(\rho^{(1\: or \:2)})$ is equal to $1$.
\end{measurefermion}

In full agreement with the previous considerations we can say that the von 
 Neumann entropy is a measure {\em} both of the amount of uncertainty deriving 
 from the indistinguishability of the involved particles {\em and} of the 
 possible uncertainty related to the entangled nature of the states (that is, 
 the impossibility
 of attributing two definite state vectors to the pair of particles).
This quantity, in the case of fermions, is strictly greater than $1$ and 
 the greater it is, the larger is the amount of
 entanglement of the state.
 
Before coming to the boson case, a last remark is appropriate. Let us 
 consider a non-entangled state of two fermions which has the form of 
 Eq.~(\ref{fermion3}).
Suppose we choose two arbitrary orthonormal vectors $\vert 
 \bar{1}\rangle$ and $\vert \bar{2}\rangle$ belonging to the two-dimensional 
 manifold spanned by the states $\vert 1\rangle$ and $\vert 2\rangle$, 
 such that:
\begin{equation}
 \label{fermion7}
 \vert 1\rangle=\alpha\vert \bar{1}\rangle+\beta\vert 
 \bar{2}\rangle\:, \hspace{2cm} \vert 2\rangle=-\beta^{*}\vert 
 \bar{1} \rangle+\alpha^{*}\vert \bar{2}\rangle\:,
\end{equation}
\noindent with $\vert \alpha\vert^{2}+ \vert \beta \vert^{2}=1$. \\
Substituting these expressions in Eq.~(\ref{fermion3}) we have:
\begin{equation}
 \label{fermion8}
 \vert \psi(1,2)\rangle = \frac{1}{\sqrt{2}}\,[\,\vert \bar{1} 
 \rangle_{1}\otimes
 \vert \bar{2}\rangle_{2} - \vert \bar{2}\rangle_{1}\otimes 
 \vert \bar{1}\rangle_{2}\,].
\end{equation}
Therefore, when a state is obtained by antisymmetrizing the product 
 of two orthogonal single-particle states, the same state can also be 
 obtained by antisymmetrizing the product of any pair of orthogonal states 
 belonging to the two-dimensional manifold spanned by the original 
 states. 
This is not surprising but it raises some questions concerning the problem 
 of the assignment of definite properties to the constituents. 
In fact, Eq.~(\ref{fermion8}) shows that, just as we can claim that in the 
 state of Eq.~(\ref{fermion3}) there is one particle with the 
 properties associated with $\vert 1\rangle$ and one with the properties 
 associated with $\vert 2\rangle$, we can make an analogous statement with 
 reference to any two arbitrary orthogonal states $\vert \bar{1}\rangle$ 
 and $\vert \bar{2}\rangle$ of the same manifold. 
However, this peculiar feature is only apparently problematic. 
In fact it seems so because here we are confining our consideration to the 
 spin degrees of freedom. 
When one takes into account also the space degrees of freedom one realizes 
 that the only physically interesting situations are those
 associated with states like the one of Eq.~(\ref{falsoepr}), that is, those 
 in which one wants, for example, to investigate the spin properties of a 
 particle that has a precise location~\footnote{Obviously, due to the 
 previous remarks concerning the arbitrariness of the states appearing 
 in Eq.~(\ref{fermion3}), also with reference to Eq.~(\ref{falsoepr}) one 
 might claim, e.g., that there is with certainty a particle in the state 
 $1/\sqrt{2} 
 [\vert z \uparrow \rangle\vert R \rangle+ \vert z \downarrow \rangle\vert 
 L\rangle]$ and one in the state $1/\sqrt{2}[\vert z \uparrow
 \rangle\vert R \rangle- \vert z \downarrow \rangle\vert  L\rangle]$.
 However such states are not physically interesting and the corresponding
 measurements are extremely difficult (if not impossible) to perform.}.
As we have seen, in such a state one can make precise claims about the spin 
 state of the particle in the Right or Left, respectively, contrary to 
 what happens for a state like the one of Eq.~(\ref{veroepr}) which is
 genuinely entangled.
This point has been exhaustively discussed in
 Ref.~\cite{gmw}, to which we refer the reader for a more detailed analysis.


\subsection{The boson case}

Let us now pass to the more delicate case of two identical bosons. In
 such a case, dealing with the problem of their entanglement requires
 great care in order to avoid  possible misunderstandings.
Let us start, as before, by considering the bosonic Schmidt
 decomposition for an arbitrary state vector $\vert \psi(1,2) \rangle$
 belonging to the symmetric manifold ${\cal S}(\mathbb{C}^{2s+1}\otimes
 \mathbb{C}^{2s+1})$ describing two identical bosons, and let us recall 
 a well-known theorem of matrix analysis
 (the so-called Takagi factorization theorem~\cite{taka}, a
 particular instance of a more general theorem named the Singular Value
 Decomposition) which is particularly useful for this case:
\begin{simcomplex}
 \label{simcomplex}
 For any symmetric $(N\times N)$ complex matrix B (that is, $B\in
 {\cal M} (N, \mathbb{C})$ and $B^{T}=B$) there exists a unitary
 transformation $U$ such that $B=U\Sigma U^{T}$, where $\Sigma$ is a real
 nonnegative diagonal matrix $\Sigma=diag[ b_{1},\dots,b_{N}]$.
The columns of $U$ are an orthonormal set of eigenvectors of
 $BB^{\dagger}$ and the diagonal entries of $\Sigma$ are the nonnegative
 square roots of the corresponding eigenvalues.
\end{simcomplex}
It is worth noticing that the columns of the unitary operator $U$ must
 be built by choosing a precise set~\footnote{With this expression 
 we mean  that one has to appropriately choose the phase factors of 
 the eigenstates.} of eigenvectors of $BB^{\dagger}$
 (to be  determined in an appropriate manner, see~\cite{taka}).
The bosonic Schmidt decomposition turns out to be a
 trivial consequence of the just mentioned Theorem~\ref{simcomplex}:
\begin{bosonschmidt}
 \label{bosonschmidt}
 Any state vector
 describing two identical $s$-spin boson particles $\vert \psi(1,2)
 \rangle$ and, consequently, belonging to the symmetric manifold
 ${\cal S}(\mathbb{C}^{2s+1}\otimes \mathbb{C}^{2s+1})$ can be written as:
\begin{equation}
 \label{boson1}
 \vert \psi(1,2)\rangle = \sum_{i=1}^{2s+1} b_{i}\,
 \vert i \rangle_{1} \otimes \vert i \rangle_{2}\:,
\end{equation}
where the states $\left\{\, \vert i \rangle \right\}$, with
 $i=1,\dots, 2s+1$, constitute an orthonormal basis for $\mathbb{C}^{2s+1}$,
 and the real nonnegative coefficients $b_{i}$ are the diagonal elements
 of the matrix $\Sigma$ and satisfy the normalization condition
 $\sum_{i}b_{i}^{2}= 1$.
\end{bosonschmidt}

The Schmidt decomposition of Eq.~(\ref{boson1}) is always unique when the 
 eigenvalues of the operator $BB^{\dagger}$ are non-degenerate, as happens
 for the biorthonormal decomposition of states describing distinguishable
 particles.

In analogy with the case of two distinguishable particles and of two
 identical fermions, one could naively  be inclined to term entangled
 any bosonic state whose Schmidt number [that is, the number of non-zero
 coefficients in the decomposition~(\ref{boson1})] is strictly greater than
 $1$ (as clearly stated, for example, in Ref.~\cite{pask}).
However, as we will show, this criterion turns out to be inappropriate because
 the states with Schmidt number equal to $1$ (that is, factorized
 states in which the two bosons are in the same state) do not exhaust the 
 class of non-entangled pairs of identical bosons.

In order to identify a satisfactory and unambiguous criterion for detecting
 entanglement in systems of identical bosons, we resort to the decomposition of
 Eq~(\ref{boson1}) and we analyze separately the cases of Schmidt number 
 equal to $1$, $2$, and greater than or equal to $3$.

\noindent{\bf Schmidt number $=1$}. In this case the factorized state
 $\vert \psi(1,2) \rangle= \vert i^{\star} \rangle \otimes \vert i^{\star}
 \rangle$ describes two identical bosons in the same state $\vert i^{\star}
 \rangle$.
It is obvious that such a state must be considered as non-entangled because 
 one knows precisely the properties of both constituents and, consequently,
 no uncertainty remains concerning which particle has which property.
This fact perfectly agrees with the von Neumann entropy of the single-particle
 reduced statistical operators $S(\rho^{(1 \,or\,2)})$ being null. \\

 \noindent{\bf Schmidt number $=2$}. According to Eq.~(\ref{boson1}), the most
 general state with Schmidt number equal to $2$ has the following form:
\begin{equation}
 \label{boson4.3}
 \vert \psi(1,2) \rangle= b_{1}\vert 1 \rangle_{1} \otimes \vert 1\rangle_{2}
 + b_{2}\vert 2 \rangle_{1} \otimes \vert 2\rangle_{2},
\end{equation}
where $b_{1}^{2}+b_{2}^{2}=1$.
The single-particle reduced density operator and its associated von Neumann 
 entropy are:
\begin{equation}
\label{boson4.35}
 \rho^{(1\,or\,2)}  =  b_{1}^{2}\,\vert 1 \rangle\langle 1\vert
 + b_{2}^{2}\,\vert 2\rangle \langle 2\vert \:,
\end{equation}
\begin{equation}
 \label{boson4.4}
 S(\rho^{(1\,or\,2)}) = - b_{1}^{2} \log b_{1}^{2}-b_{2}^{2}\log b_{2}^{2}.
\end{equation}
We can now distinguish two cases, depending on the values of the nonnegative 
 coefficients $b_{1}$ and $b_{2}$. 

First we consider the case where the two coefficients are equal, that is, 
 $b_{1}= b_{2}=1/\sqrt{2}$, and we prove the following Theorem:
\begin{bequalb}
 \label{bequalb}
The condition $b_{1}=b_{2}=1/\sqrt{2}$ is necessary and sufficient in order
 that the state $\vert \psi(1,2) \rangle= b_{1}\vert 1 \rangle_{1} \otimes
 \vert 1\rangle_{2} + b_{2}\vert 2 \rangle_{1} \otimes \vert 2\rangle_{2}$
 might be obtained by symmetrizing the factorized product of two orthogonal
 states.
\end{bequalb}
{\em Proof.} Suppose that the state $\vert \psi(1,2)\rangle$ is obtained by
 symmetrizing the tensor product of two orthogonal states $\vert \phi \rangle$
 and $\vert \chi \rangle$.
If we define the orthogonal states 
 $\vert 1\rangle=1/\sqrt{2}\,(\vert \phi \rangle + \vert
 \chi \rangle)$ and $\vert 2\rangle=i/\sqrt{2}\,(\vert \phi \rangle - \vert
 \chi \rangle)$, we get:
\begin{equation}
 \label{boson4.41}
 \vert \psi(1,2)\rangle= \frac{1}{\sqrt{2}}\,[\,\vert \phi\rangle_{1} \otimes
 \vert \chi \rangle_{2} + \vert \chi \rangle_{1} \otimes \vert \phi 
 \rangle_{2}\,] = \frac{1}{\sqrt{2}}\,[\,\vert 1 \rangle_{1} \otimes
 \vert 1 \rangle_{2} + \vert 2 \rangle_{1} \otimes \vert 2  \rangle_{2}\,].
\end{equation}
On the contrary, if $b_{1}\!=\!b_{2}\!=\!1/\sqrt{2}$ in
 Eq.~(\ref{boson4.3}), then by defining the orthogonal
 states $\vert \phi \rangle=1/\sqrt{2}\,(\vert 1 \rangle -i\vert
 2\rangle)$ and $\vert \chi \rangle=1/\sqrt{2}\,(\vert 1 \rangle +i \vert
 2 \rangle)$, one gets:
\begin{equation}
 \label{boson4.42}
 \vert \psi(1,2)\rangle= \frac{1}{\sqrt{2}}\,[\,\vert \phi\rangle_{1} \otimes
 \vert \chi \rangle_{2} + \vert \chi \rangle_{1} \otimes \vert \phi 
 \rangle_{2}\,],
\end{equation}
thus showing that $\vert \psi(1,2) \rangle$ can be obtained by 
 symmetrizing the factorized product of two orthogonal states $\blacksquare$. 
 
In this situation, and in full accordance with our
 Theorem~\ref{factorizabilityidentical3}, one must consider this state as a
 non-entangled one since it is possible to
 attribute definite state vectors to both particles (of course, once
 again, we cannot say which particle is associated with the state
 $\vert \phi \rangle$ or $\vert \chi \rangle$ because of their
 indistinguishability). 
Actually, such a state has precisely {\it the same conceptual status and
 exhibits the same physical features} as the non-entangled states of a pair of
 fermions and the reduced density operator $\rho^{(1\,or\,2)}$ 
 describes the physical state of one randomly chosen boson (given
 that one of them is associated with the state $\vert \phi \rangle$
 while the other to the state $\vert \chi \rangle$):
\begin{equation}
\label{boson4.421}
 \rho^{(1\,or\,2)} = \frac{1}{2}\,[\,\vert \phi \rangle \langle \phi \vert
 + \vert \chi \rangle \langle \chi\vert \,].
\end{equation}
Moreover, since the state contains no correlations but those descending
 from the exchange of the two identical particles, the von Neumann
 entropy $S(\rho^{(1\,or\,2)})$, according to Eq.~(\ref{boson4.4}), is
 correctly equal to $1$ and our ignorance concerns only which particle has to
 be associated with which state.

Before proceeding a short digression is appropriate. 
First, if one chooses any pair of orthonormal vectors that are a linear
 combination, {\em with real coefficients}, of the above states
 $\vert 1\rangle$ and $\vert 2\rangle$:
\begin{equation}
 \label{boson3.1}
 \vert \bar{1}\rangle = \alpha \vert 1\rangle + \beta \vert 2\rangle\:,
 \hspace{2cm} \vert 
 \bar{2} \rangle = -\beta \vert 1\rangle + \alpha\vert 2 \rangle\:,
\end{equation}
where $\alpha,\beta \in \mathbb{R}$ and $\alpha^{2}+\beta^{2}=1$, one has:
\begin{equation}
 \label{boson3.2}
 \vert\psi (1,2)\rangle=\frac{1}{\sqrt{2}}\,[\,\vert 1\rangle_{1}\otimes
 \vert 1 \rangle_{2}+\vert 2\rangle_{1} \otimes  \vert 2\rangle_{2}\,] =
 \frac{1}{\sqrt{2}}\,[\,\vert \bar{1}\rangle_{1}\otimes
 \vert  \bar{1}\rangle_{2}+\vert \bar{2}\rangle_{1} \otimes 
 \vert \bar{2}\rangle_{2}\,]\:,
\end{equation}
Thus, as in the case of distinguishable particles, when degeneracy occurs, 
 the Schmidt decomposition is not unique.

This situation recalls the one we met in the case of two 
 non-entangled fermions. 
However, the possibility of writing $\vert\psi (1,2)\rangle$ in
 different Schmidt forms, contrary to what happens with fermions, is 
 absolutely unproblematic from the physically interesting point of 
 view of the properties possessed by the two particles. 
Actually, the pair of orthonormal states whose symmetrization leads to 
 the expression for the state of Eq.~(\ref{boson4.42}) is uniquely determined
 in the present case. 
Accordingly, for identical bosons there is no ambiguity 
 concerning the properties which allow us to claim that in the state
 of Eq.~(\ref{boson4.42}) {\em ``there is one particle with the properties 
 identified by $\vert\phi\rangle$ and one with the properties 
 identified by $\vert\chi\rangle$"}.
It is illuminating to stress the difference between two ways of 
 looking at the properties of the constituents according to what 
 emerges naturally when one expresses the state in the form of 
 Eq.~(\ref{boson4.42}) or of Eq.~(\ref{boson3.2}).
Actually, if one looks at the form of Eq.~(\ref{boson3.2}), without 
 taking into account that it has been obtained by symmetrizing a factorized 
 pair of orthogonal states, one is naturally led to make different statements, 
 which however have a {\em probabilistic} character, i.e., to assert that 
 {\em ``with probability $1/2$ both particle are (in the sense they will be 
 found to be) in state $\vert 1\rangle$ (or $\vert \bar{1}\rangle$) and, 
 with the same probability they are in state $\vert 2\rangle$
 (or $\vert \bar{2} \rangle$)"}. 

Let us now consider the second case, i.e., the one in which $b_{1}\neq b_{2}$,
 and prove the following Theorem:
\begin{bdifferentb}
 \label{bdifferentb}
The condition $b_{1}\neq b_{2}$ is necessary and sufficient in order that
 the state $\vert \psi(1,2) \rangle= b_{1}\vert 1 \rangle_{1} \otimes
 \vert 1\rangle_{2} + b_{2}\vert 2 \rangle_{1} \otimes \vert 2\rangle_{2}$
 might be obtained by symmetrizing the factorized product of two
 non-orthogonal states.
\end{bdifferentb}
{\em Proof.} Since the Schmidt number of $\vert \psi(1,2)\rangle$, and as
 a consequence the rank of the reduced statistical operators, is equal to
 $2$, the state is obtained either by symmetrizing the factorized product 
 of two orthogonal
 states or by symmetrizing the product of two non-orthogonal states. 
In fact, if at least three linearly independent states were involved in the
 symmetrization procedure, the rank of the single-particle statistical operator
 would consequently be strictly greater than $2$.
This Theorem is then simply the logical negation of the previous
 Theorem~\ref{bequalb} $\blacksquare$. 

As we have just proved, in this case the state $\vert \psi(1,2)\rangle$ can
 be obtained by symmetrizing a factorized product of two non-orthogonal
 states $\vert \phi \rangle$ and $\vert \chi \rangle$:
\begin{equation}
 \label{boson3.35}
 \vert \psi(1,2)\rangle = \frac{1}{\sqrt{2(1+\vert \langle \chi \vert
 \phi \rangle \vert^{2})}}\,[\,
 \vert  \phi \rangle_{1}\otimes \vert  \chi \rangle_{2} +
 \vert \chi \rangle_{1}\otimes \vert \phi\rangle_{2}\,],
 \hspace{1cm} \langle \chi \vert \phi \rangle \neq 0.
\end{equation}
Then, if one defines the vector $\vert \phi_{\perp}\rangle$
 as the unique normalized vector orthogonal to $\vert \phi \rangle$ and
 lying in the two-dimensional manifold spanned by $\vert \phi \rangle$ and
 $\vert \chi \rangle$, the state of Eq.~(\ref{boson3.35}) can be written
 as~\footnote{We notice that one could have equivalently defined the
 vector $\vert \chi_{\perp} \rangle$ as the unique normalized vector orthogonal
 to $\vert \chi \rangle$ and lying in the two-dimensional manifold spanned by
 $\vert \phi \rangle$ and $\vert \chi \rangle$, without modifying the
 forthcoming conclusions.}:
\begin{equation}
 \label{boson4}
 \vert \psi(1,2)\rangle =
 a\vert \phi \rangle_{1} \otimes \vert \phi\rangle_{2} +
 \frac{b}{\sqrt{2}}\,[\,
 \vert \phi \rangle_{1}\otimes \vert \phi_{\perp} \rangle_{2} +
 \vert \phi_{\perp} \rangle_{1}\otimes \vert \phi \rangle_{2}\,]\,,
\end{equation}
\noindent with certain complex coefficients $a,b\neq 0$ satisfying the
 normalization condition $\vert a \vert^{2}+ \vert b \vert^{2} =1$ and 
 depending on the modulus of the scalar product $\vert \langle \chi
 \vert \phi \rangle \vert$ in the following way: $\vert b\vert =
 (1- \vert \langle \chi \vert \phi \rangle \vert^{2})^{1/2}\cdot
 (1+\vert \langle \chi \vert \phi \rangle \vert^{2})^{-1/2}$.

In this case, the criteria adopted in Refs.~\cite{pask,li,vacca}
 correctly lead one to declare the state of Eq.~(\ref{boson4}) as a genuine 
 entangled one.
Since this state is obtained by symmetrizing a factorized product of 
 two {\it non-orthogonal} vectors, it is impossible to attribute to both 
 particles definite properties, so that it has to be considered as 
 entangled within our approach also.
It is of some interest to exhibit explicitly the Schmidt decomposition 
 of the state of Eq.~(\ref{boson4}):
\begin{equation}
 \label{boson4.1}
 \vert \psi(1,2)\rangle =
 \sqrt{\frac{1+\sqrt{1-\vert b\vert^{4}}}{2}}
 \vert 1 \rangle_{1} \otimes \vert 1\rangle_{2} +
 \sqrt{\frac{1-\sqrt{1- \vert b\vert^{4}}}{2}}
 \vert 2 \rangle_{1} \otimes \vert 2\rangle_{2}
 \hspace{0.8cm}  \vert b \vert \in (0,1)\,,
\end{equation}
where the orthonormal states $\vert 1 \rangle$ and $\vert 2 \rangle$ are
 defined as:
\[
 \vert 1 \rangle =
 \frac{\vert b\vert e^{i\theta_{a}/2}}
 {\sqrt{2-\vert a \vert^{2}- \sqrt{1- \vert b \vert^{4}}}}
 \Big[ \:\vert \phi \rangle +
 \frac{\sqrt{1-\vert b\vert^{4}} - \vert a \vert^{2}}{\sqrt{2}ab^{\star}}
 \vert \phi_{\perp} \rangle \:\Big]\,,
\]
\begin{equation}
 \label{boson4.2}
 \vert 2 \rangle =
 \frac{ i\vert b\vert e^{i\theta_{a}/2}}
 {\sqrt{2-\vert a \vert^{2}+ \sqrt{1- \vert b \vert^{4}}}}
 \Big[ \:\vert \phi \rangle -
 \frac{\sqrt{1-\vert b\vert^{4}}+ \vert a \vert^{2}}{\sqrt{2}ab^{\star}}
 \vert \phi_{\perp} \rangle \:\Big]\,,
\end{equation}
with $a=\vert a\vert e^{i\theta_{a}}$, $a$ and
 $b\neq 0$, and $\vert a \vert^{2} + \vert b \vert^{2}=1$.
It is worth noticing that, contrary to what happens for the non-entangled 
 state of Eq.~(\ref{boson4.42}), the Schmidt decomposition of 
 Eq.~(\ref{boson4}) is now uniquely determined.
This reflects the fact that the two eigenvalues of the operator $BB^{\dagger}$,
 where $B$ is the matrix of the coefficients of the decomposition
 of $\vert\psi(1,2)\rangle$ of Eq.~(\ref{boson4}) on a factorized single
 particle basis including $\vert \phi \rangle$ and $\vert \phi_{\perp}
 \rangle$, are distinct.

Let us come now to a consideration, for the case under investigation, of
 the reduced density operator describing one randomly chosen particle and 
 of its associated von Neumann entropy.
They are given by Eqs.~(\ref{boson4.35}) and~(\ref{boson4.4}) which immediately
 show that the entropy belongs to the open interval $(0,1)$.
Actually, assuming that $b_{1}^{2}>1/2$, in a measurement process there is a
 probability greater than $1/2$ of finding both bosons in the 
 same physical state $\vert 1\rangle$, and a probability less than $1/2$ of 
 finding them both in the orthogonal state $\vert 2 \rangle$. 
Accordingly, in this situation we have more information about the single 
 particle state than in the case of the non-entangled state of 
 Eq.~(\ref{boson4.42}) and, as a consequence, the von Neumann entropy is
 strictly less than $1$. \\

\noindent {\bf Schmidt number $\geq 3$}. In this situation, the state is a
 genuine entangled one since
 it cannot be obtained by symmetrizing a factorized product of two
 orthogonal states~\footnote{In fact, if this was true, the rank of the
 reduced density operator would be equal to $2$, in contradiction with the
 fact that a Schmidt number greater than or equal to $3$ implies a rank equal
 to or greater than $3$.} and the von Neumann entropy
 of the reduced density operators is such that $S(\rho^{(1\,or\,2)})
 \in (0, \log(2s+1)]$.
As before, the entropy is close to zero when, in the Schmidt
 decomposition of the state, one of its coefficients is very close to $1$, 
 implying that there is a very high probability of being right in claiming
 that both bosons are (in the usual quantum sense of {\em ``will be found to
 be if a measurement is performed"}) in the state associated with the largest
 coefficient.
On the contrary, the entropy equals $\log(2s+1)$ when the
 decomposition involves all the basis states of $\mathbb{C}^{2s+1}$
 with equal weights (a maximally entangled state).
Moreover, as we will prove below, when the entropy lies within the interval
 $(0,1)$, there is a 
 precise state such that the probability of finding a (randomly chosen) 
 particle in it, in the appropriate measurement, is greater than $1/2$.
Obviously, the identification of the privileged state mentioned requires a
 knowledge of the Schmidt decomposition, since this state is the one associated
 to the largest $b_{i}$ appearing in Eq.~(\ref{boson1}).
This feature is a consequence of the following Theorem:
\begin{entropyin01}
 \label{entropyin01}
 Consider the statistical operator $\rho = \sum_{i} \,p_{i} \vert i \rangle
 \langle i\vert$ where at least three of its weights $p_{i}$ are different
 from zero. If each non-zero weight $p_{i}$ is strictly less than $1/2$,
 then the von Neumann entropy $S(\rho)$ is strictly greater than $1$. 
\end{entropyin01}
{\em Proof.} Let us consider the following function $f(p_{1},p_{2})$ of
 two variables:
\begin{equation}
 \label{boson4.51}
 f(p_{1},p_{2}) \equiv -\sum_{i=1}^{2}p_{i}\,\log p_{i} -(1-p_{1}-p_{2})\,\log
 (1-p_{1}-p_{2}) 
\end{equation}
with $p_{1},p_{2}\in (0,1/2)$ and $p_{1}+p_{2}>1/2$.
The only stationary point of $f(p_{1},p_{2})$ in the considered (open) region
 is a maximum, while the minimum value of $f$ is attained on the boundary of
 such a region and it equals $1$.
As a consequence, given a rank-$3$ statistical operator $\rho = \sum_{i=1}^{3}
 p_{i}\vert i \rangle \langle i \vert$, its von Neumann entropy $S(\rho)
 = f(p_{1},p_{2})$ is strictly greater than $1$ whenever
 $p_{i}\in (0,1/2),\:i=1,2,3$.
Let us now suppose that the Theorem holds true for an arbitrary rank $n>3$
 statistical operator $\rho= \sum_{i=1}^{n} p_{i}\vert i \rangle \langle i
 \vert$ and let us prove that it also holds true for an arbitrary rank $n+1$
 statistical operator in the following way. 
Take any
 $p_{1}$, which by assumption is smaller than $1/2$, and split it into two
 positive weights $p_{11}$ and $p_{12}$, so that $p_{1}=p_{11}+p_{12}$.
A basic property of the Shannon entropy function $-\sum_{i} p_{i} \log p_{i}$
 tells us that:
\begin{equation}
 \label{boson4.52}
 -(p_{11}+p_{12})\log (p_{11}+p_{12}) - \sum_{i=2}^{n} p_{i}\log p_{i} <
  -p_{11}\log p_{11} - p_{12}\log p_{12}- \sum_{i=2}^{n} p_{i}\log p_{i}
\end{equation}
By hypothesis the left hand side
 of Eq.~(\ref{boson4.52}) is strictly greater than $1$
 since it is the von Neumann entropy of a statistical operator of rank $n$ with
 weights strictly less than $1/2$, while the RHS is the von Neumann
 entropy of a statistical operator of rank $n+1$.
Since the weights $(p_{11},p_{12},p_{2},\ldots,p_{n})$, apart from being
 constrained to belong to the interval $(0,1/2)$, are totally arbitrary,
 we have proved that the von Neumann entropy of any rank-($n+1$) statistical
 operator, with all weights belonging to $(0,1/2)$, is strictly greater
 than $1$.
By induction the Theorem holds true for all $n>3$ $\blacksquare$.\\

Summarizing the previous analysis, we have seen that, in the case of two
 identical bosons, consideration of the Schmidt number alone 
 to detect the entanglement of a state fails since there exist bosonic states
 with Schmidt number equal to $2$ which can be entangled as well as
 non-entangled.
Similarly, we have seen that the von Neumann entropy of the reduced statistical
 operators of the non-entangled state of Eq.~(\ref{boson4.41}) is equal to
 $1$, and that the same value can characterize entangled states with Schmidt
 number greater than or equal to $3$.
Accordingly, the von Neumann entropy criterion also does not allow, by
 itself, an unambiguous identification of non-entangled states.

Concluding, we have shown that a complete and satisfactory criterion for
 distinguishing entangled from non-entangled boson states should involve
 consideration of {\em both} the Schmidt number criterion {\em and} the von 
 Neumann
 entropy of the reduced density operator of each single-particle, in accordance
 with the following Theorem:
\begin{measureboson}
 \label{measureboson}
 A state vector $\vert \psi(1,2) \rangle$ describing two identical
 bosons is {\bf non-entangled} if and only if either its Schmidt number is 
 equal to $1$, or the Schmidt number is equal to $2$ {\em and}  the 
 von Neumann entropy of
 the one-particle reduced density operator $S(\rho^{(1\: or \:2)})$ is
 equal to $1$. 
Alternatively, one might say that the state is {\bf 
 non-entangled} if and only if either its von Neumann entropy is equal to $0$, 
 or it is equal to $1$ and the Schmidt number is equal to $2$.
\end{measureboson}
%


\section{Summary}

To conclude our paper we  summarize  the previous results
 and give a general overview of the problem of characterizing the
 entanglement of quantum systems composed of two identical particles. 
First of all we recall that this problem can be dealt with in a completely 
 rigorous way by following our original criterion of Ref.~\cite{gmw}
 expressed in the Definition~\ref{defofentangidentical}. 
However, here we are interested in analyzing the criteria based on the 
 determination of the Slater and the Schmidt numbers of the
 (fermionic and bosonic, respectively)
 Schmidt decomposition of the  state vector of the composite system and on 
 the evaluation of the von Neumann entropy of the reduced density operators 
 associated with each single constituent. 
As usual, we deal separately with the cases of two 
 identical fermions and of two identical bosons.

We start with a system of two identical fermions. \\

\noindent {\bf Two identical fermions}
\begin{enumerate}
\item Slater number of $\vert \psi(1,2) \rangle = 1  \hspace{0.2cm}
\Leftrightarrow \hspace{0.2cm} S(\rho^{(1\,or\,2)})=1
\hspace{0.2cm}\Leftrightarrow$ \hspace{0.2cm} non-entangled state
\item Slater number of $\vert \psi(1,2) \rangle > 1 \hspace{0.2cm}
\Leftrightarrow \hspace{0.2cm} S(\rho^{(1\,or\,2)})> 1
\hspace{0.2cm} \Leftrightarrow$ \hspace{0.2cm} entangled state
\end{enumerate}
In the first case the state can be obtained by antisymmetrizing a tensor
 product of two orthogonal single-particle states, while in the second
 case it cannot.

In the considered case the two criteria are, as indicated, fully 
 equivalent, and they identify precisely the same states as our 
 criterion.
It is interesting to notice that, since identical fermions cannot be in the 
 same state, at least the uncertainty corresponding to the impossibility of 
 identifying which particle has which property, when there is no entanglement,
 is always present.
This is why the minimum value of the entropy turns out to be equal to 1. 
In such a case, when one makes a precise claim concerning the state of a 
 {\em randomly chosen particle}, there is a probability
 equal to $1/2$ that the claim is correct. 
On the other hand, when the Slater number is greater than 1, or, equivalently,
 the entropy is larger than 1, there is no state such that the statement 
 {\em ``this particle, if subjected to the appropriate measurement, will be 
 found in such a state"} has a probability equal to or greater than $1/2$. 
Thus, the non-entangled state is the one in which one has 
 the maximum information about the state of the system.

The case of two identical bosons is slightly more involved. We 
 recall the conclusions we have reached:\\

\noindent {\bf Two identical bosons}
 \begin{enumerate}
 \item Schmidt number of $\vert \psi(1,2) \rangle = 1 \hspace{0.2cm}
 \Leftrightarrow \hspace{0.2cm} S(\rho^{(1\,or\,2)})=0
\hspace{0.2cm} \Rightarrow$ \hspace{0.2cm} non-entangled state
 \item Schmidt number of $\vert \psi(1,2) \rangle = 2$ and
 $S(\rho^{(1\,or\,2)})\in (0,1)
 \hspace{0.2cm} \Rightarrow$\hspace{0.2cm} entangled state
 \item Schmidt number of $\vert \psi(1,2) \rangle = 2$ and
 $S(\rho^{(1\,or\,2)})=1
 \hspace{0.2cm} \Rightarrow$\hspace{0.2cm} non-entangled state
 \item Schmidt number of $\vert \psi(1,2) \rangle > 2$ \hspace{0.2cm}
 $\Rightarrow$ \hspace{0.2cm} entangled state
\end{enumerate}
In the first case the state is the tensor product of the same single
 particle vector; in the second case the state can be obtained by
 symmetrizing two non-orthogonal vectors and
 thus no definite property can be attributed to both subsystems.
In the third case the state can be obtained by symmetrizing a tensor
 product of two orthogonal vectors while in the last one it 
 involves more than two linearly independent single-particle states and 
 the von Neumann entropy $S(\rho^{(1\,or\,2)})$ can take any value
 within the interval $(0,\log (2s+1)]$.
The above list exhibits some interesting features. First of all it 
 shows clearly that the Schmidt number cannot be used, unless it has 
 the value 1, to identify non-entangled states, since there are both entangled 
 and non-entangled states with  Schmidt number equal to 2. 
It also shows that the von Neumann entropy criterion does not allow, 
 by itself, a clear-cut identification of the non-entangled states since 
 it can take the value 1 both for a non-entangled state of
 Schmidt number 2 and for an entangled state of Schmidt number greater than 2.

As discussed in this paper, the von Neumann entropy supplies us, by 
 itself, with important information concerning the state of a randomly 
 chosen constituent: when it lies in the interval
 $(0,1)$ there is a precise state such that the probability of 
 finding it in an appropriate measurement is greater than $1/2$.


\section{Conclusions}

In this paper we have discussed in general the problem of
 deciding whether a state describing a system of two identical
 particles is entangled or not. 
We recalled the general criteria and results derived in 
 Refs.~\cite{gmw,gm}, which make explicit reference to the possibility of 
 attributing a complete set of objective properties to both 
 constituents. 
We have shown how this attitude allows one to clarify the somewhat puzzling 
 situation that one meets when resorting  to the consideration either of the
 Slater and the Schmidt number of the (fermionic and bosonic) Schmidt
 decomposition or of the von Neumann 
 entropy of the reduced statistical operator for detecting 
 entanglement. 
Our analysis has made clear that some alleged difficulties of the
 mentioned approaches derive simply from not having appropriately taken into 
 account the peculiar role of the identity within the quantum formalism.
In particular, we have shown how the consideration of the von 
 Neumann entropy allows one to determine whether the  uncertainty concerning 
 the states arises simply from the identity of the particles or is also a 
 genuine consequence of the entanglement.
With reference to this problem we stress once more that, when all 
 our lack of information derives simply
 from the identity, the existing correlations cannot be used
 as a quantum mechanical resource to implement any teleportation
 procedure or to violate Bell's inequality, contrary to what happens
 for an entangled state.


\section{Acknowledgements}
We thank the Referee for his lucid and useful comments and remarks.




\begin{thebibliography}{99}

\bibitem{sch} E.Schr\"odinger, {\it Proceedings of the Cambridge Philosophical 
 Society} {\bf 31}, 555 (1935); {\bf 32}, 446 (1936). 

\bibitem{gmw} G.C.Ghirardi, L.Marinatto and T.Weber, {\it J. Stat. Phys.} 
 {\bf 108}, 49 (2002).

\bibitem{gm} G.C.Ghirardi and L.Marinatto, {\it Fortschr. Phys.} {\bf 51},
 No.4-5, 379 (2003).

\bibitem{epr} A.Einstein, B.Podolsky and N.Rosen,  {\it Phys. Rev.}
 {\bf 47}, 777 (1935).

\bibitem{cirac} J.Schliemann, J.I.Cirac, M.Kus, M.Lewenstein, and D.Loss,
 {\it Phys. Rev. A} {\bf 64}, 022303 (2001).

\bibitem{pask} R.Paskauskas and L.You, {\it Phys. Rev. A} {\bf 64},
 042310 (2001).

\bibitem{li} Y.S.Li, B.Zeng, X.S.Liu and G.L.Long, {\it Phys. Rev. A}
 {\bf 64}, 054302 (2001).

\bibitem{vacca} H.M.Wiseman and J.A.Vaccaro, {\it Phys. Rev. Lett.}
 {\bf 91}, 097902 (2003).

\bibitem{ghsz} J.F.Clauser, M.A.Horne, A.Shimony and R.A.Holt,
 {\it Phys. Rev. Lett.} {\bf 26}, 880 (1969).

\bibitem{mehta} M.L.Mehta, {\it Matrix Theory: Selected Topics and Useful
 Results}, Les Editions de Physique, Les Ulis cedex, France (1989).

\bibitem{taka} R.A.Horn and C.R.Johnson, {\it Matrix Analysis}, Cambridge
 University Press, Cambridge, England (1986).

\end{thebibliography}
 \end{document}